\documentclass[aps,pra,reprint,superscriptaddress,showpacs, amsmath,amssymb]{revtex4-1}
\usepackage{graphicx}
\usepackage[pdfstartview=FitH,bookmarks=false,colorlinks=true, linkcolor=blue,citecolor=blue,urlcolor=blue]{hyperref}

\begin{document}

\title{Spin correlation tensor for measurement of quantum entanglement \\
in electron-electron scattering}
\author{D. E. Tsurikov}	\email[]{DavydTsurikov@mail.ru}
\affiliation{Institute of Physics, St. Petersburg State University, St. Petersburg 199034, Russia}
\author{S. N. Samarin}
\affiliation{Institute of Physics, St. Petersburg State University, St. Petersburg 199034, Russia}
\affiliation{School of Physics, the University of Western Australia, Crawley, Perth 6009, WA, Australia}
\author{J. F. Williams}
\affiliation{School of Physics, the University of Western Australia, Crawley, Perth 6009, WA, Australia}
\author{O. M. Artamonov}
\affiliation{Institute of Physics, St. Petersburg State University, St. Petersburg 199034, Russia}

\date{\today}

\begin{abstract}
We consider the problem of correct measurement of a quantum entanglement in the two-body electron-electron scattering. An expression is derived for a spin correlation tensor of a pure two-electron state. A geometrical measure of a quantum entanglement as the distance between two forms of this tensor in entangled and separable cases is presented. We prove that this measure satisfies properties of a valid entanglement measure: nonnegativity, discriminance, normalization, non-growth under local operations and classical communication. This measure is calculated for a problem of electron-electron scattering. We prove that it does not depend on the azimuthal rotation angle of the second electron spin relative to the first electron spin before scattering. Finally, we specify how to find a spin correlation tensor and the related measure of a quantum entanglement in an experiment with electron-electron scattering.
\end{abstract}



\maketitle

\section{Introduction}

Quantum entanglement is of interest to modern physics, both from fundamental and applied points of view. The applied aspect of studying of an entanglement is related to its application in quantum information technology. The fundamental interest is related to the violation of the principle of locality in quantum mechanics. This principle was formulated originally in the form of the Einstein--Podolsky--Rosen paradox \cite{Bib01} and later as Bell's theorem \cite{Bib02, Bib03}. The violation of the Bell's inequalities, on which the theorem is based, was the first way of the identification of the quantum entanglement.

Now few criteria are developed \cite{Bib04, Bib05} for the identification of quantum entanglement in a system. However, they do not give the quantitative information about it. Measures of a quantum entanglement serve for this purpose \cite{Bib05}. They have to satisfy a number of requirements \cite{Bib04}. In a two-particle system, entropy satisfies all main requirements. To calculate entropy, it is necessary to find a density matrix of system. For measurement of a density matrix, the method of a quantum tomography is used \cite{Bib06}. However, a quantum tomography has not yet been performed for many problems. One of them is the two-body electron-electron scattering. Therefore, the search for a method of quantum entanglement measurement, which can be realized in the scattering experiment, is still desirable.

An approach based on the norm of a spin correlation tensor for the measurement of quantum entanglement of a system is well known \cite{Bib07, Bib08}. The advantage of this geometrical measure of entanglement is that it can be measured experimentally. However, this approach in its current form is not obvious.

In this work, we will present an obvious geometrical measure of a quantum entanglement based on a spin correlation tensor in electron-electron scattering. For that purpose we consider two-electron system in a state of coherent superposition of pairs of one-electron states \cite{Bib09}:
\begin{eqnarray}	\label{Eq01}
	| \psi \rangle = N ( && c_{++}{| + \rangle}_1{| + \rangle}_2 + c_{+-}{| + \rangle}_1{| - \rangle}_2 \nonumber\\
	&& + c_{-+}{| - \rangle}_1{| + \rangle}_2 + c_{--}{| - \rangle}_1{| - \rangle}_2 ).
\end{eqnarray}
In expression (\ref{Eq01}) ${| + \rangle}_a$ and ${| - \rangle}_a$ are orthonormal states of $a$th electron with a spin "up" and "down" respectively concerning the allocated direction:
\begin{equation}	\label{Eq02}
	{}_a{\langle + | + \rangle}_b = I_{ab} = {}_a{\langle - | - \rangle}_b, \ {}_a{\langle + | - \rangle}_b = 0 = {}_a{\langle - | + \rangle}_b,
\end{equation}
where $I$ is the unit matrix. Here, and everywhere below, the indexes $a,b\in \{1,2\}$. Taking into account a normalization $\langle \psi | \psi \rangle = 1$ for a state (\ref{Eq01}) we have
\begin{equation}	\label{Eq03}
	N = \left( {| c_{++} |}^2 + {| c_{-+} |}^2 + {| c_{+-} |}^2 + {| c_{--} |}^2 \right)^{-1/2}.
\end{equation}
For the system in the state (\ref{Eq01}), we will obtain an expression for the spin correlation tensor. We will accept the norm of the difference of this tensor and the tensor product of the electron polarization vectors as a measure of a quantum entanglement and we will prove that it is valid. We will calculate this quantity for the problem of Coulomb electron-electron scattering \cite{Bib10}. Finally, we will specify how to find the spin correlation tensor and the related measure of the quantum entanglement in the scattering experiment.

\section{Spin correlation tensor}

For the description of quantum correlations in the two-electron system the \textit{spin correlation tensor} is used \cite{Bib09, Bib11}:
\begin{equation}	\label{Eq04}
	T_{ij} := \langle {\sigma}_{i1}{\sigma}_{j2} \rangle.
\end{equation}
The mean value of the physical quantity $A$ is written in terms of a wave function
\begin{equation}	\label{Eq05}
	\langle A \rangle = \langle \psi |A| \psi \rangle
\end{equation}
or density matrix $\rho := | \psi \rangle \langle \psi |$
\begin{equation}	\label{Eq06}
	\langle A \rangle = \operatorname{Tr}\left( \rho A \right),
\end{equation}
where $\operatorname{Tr}$ is a trace on pairs of one-electron states ${{| \pm \rangle}_1}{{| \pm \rangle}_2}$ and ${{| \pm \rangle}_1}{{| \mp \rangle}_2}$. Here and everywhere below the indexes $i,j,k,l\in \{x,y,z\}$, with summation on the repeating indexes. We write the dimensionless projection of $a$th electron spin on coordinate axis $i$ as an operator:
\begin{eqnarray}	\label{Eq07}
	{\sigma}_{ia} = && {{| + \rangle}_a}\sigma _i^{11}{}_a\langle + |+{{| + \rangle}_a}\sigma _i^{12}{}_a\langle - | \nonumber\\
		&& + {{| - \rangle}_a}\sigma _i^{21}{}_a\langle + |+{{| - \rangle}_a}\sigma _i^{22}{}_a\langle - |,
\end{eqnarray}
where $\sigma _i^{ab}$ are elements of the Pauli matrices
\begin{equation}	\label{Eq08}
	{\sigma}_x :=
	\left( \begin{matrix}
		0 & 1 \\
		1 & 0 \\
	\end{matrix} \right), \
	{\sigma}_y :=
	\left( \begin{matrix}
		0 & -i \\
		i & 0 \\
	\end{matrix} \right), \
	{\sigma}_z :=
	\left( \begin{matrix}
		1 & 0 \\
		0 & -1 \\
	\end{matrix} \right).
\end{equation}

\subsection{Elements of tensor}

Let's find elements of the tensor (\ref{Eq04}) taking into account the definition (\ref{Eq05}) for the system in state (\ref{Eq01}). From the expressions (\ref{Eq01}) and (\ref{Eq07}) taking into account the property (\ref{Eq02}) we have
\begin{eqnarray}	\label{Eq09}
	\langle \psi | {\sigma}_{i1} && N^{-1} \nonumber\\
		= && \left( \sigma _i^{11}{}_1\langle + |+\sigma _i^{12}{}_1\langle - | \right)\left( {{{\bar{c}}}_{++}}{}_2\langle + |+{{{\bar{c}}}_{+-}}{}_2\langle - | \right) \nonumber\\
			&& +  \left( \sigma _i^{21}{}_1\langle + |+\sigma _i^{22}{}_1\langle - | \right)\left( {{{\bar{c}}}_{-+}}{}_2\langle + |+{{{\bar{c}}}_{--}}{}_2\langle - | \right), \nonumber\\
\end{eqnarray}
\begin{eqnarray}	\label{Eq10}
	{\sigma}_{j2} | \psi \rangle && N^{-1} \nonumber\\
		= &&  \left( {{| + \rangle}_2}\sigma _j^{11}+{{| - \rangle}_2}\sigma _j^{21} \right)\left( c_{++}{{| + \rangle}_1}+c_{-+}{{| - \rangle}_1} \right) \nonumber\\
			&& +  \left( {{| + \rangle}_2}\sigma _j^{12}+{{| - \rangle}_2}\sigma _j^{22} \right)\left( c_{+-}{{| + \rangle}_1}+c_{--}{{| - \rangle}_1} \right). \nonumber\\
\end{eqnarray}
Combining expressions (\ref{Eq09}) and (\ref{Eq10}) in the definition (\ref{Eq04}) we derive
\begin{eqnarray}	\label{Eq11}
	T_{ij} N && ^{-2} \nonumber\\
		= &&  \left( \sigma _i^{11}c_{++}+\sigma _i^{12}c_{-+} \right)\left( {\bar{c}_{++}}\sigma _j^{11}+{\bar{c}_{+-}}\sigma _j^{21} \right) \nonumber\\
			&& + \left( \sigma _i^{11}c_{+-}+\sigma _i^{12}c_{--} \right)\left( {\bar{c}_{++}}\sigma _j^{12}+{\bar{c}_{+-}}\sigma _j^{22} \right) \nonumber\\
			&& + \left( \sigma _i^{21}c_{++}+\sigma _i^{22}c_{-+} \right)\left( {\bar{c}_{-+}}\sigma _j^{11}+{\bar{c}_{--}}\sigma _j^{21} \right) \nonumber\\
			&& + \left( \sigma _i^{21}c_{+-}+\sigma _i^{22}c_{--} \right)\left( {\bar{c}_{-+}}\sigma _j^{12}+{\bar{c}_{--}}\sigma _j^{22} \right).
\end{eqnarray}
Taking into account definitions of the Pauli matrices (\ref{Eq08}) from expression (\ref{Eq11}) we have for the rows of the tensor~(\ref{Eq04}):
\begin{eqnarray}	\label{Eq12}
	T_{1j}N && ^{-2} \nonumber\\
		= &&  \sigma _j^{11}2\operatorname{Re}\left( c_{-+}{\bar{c}_{++}} \right)+\sigma _j^{12}\left( c_{--}{\bar{c}_{++}}+c_{+-}{\bar{c}_{-+}} \right) \nonumber\\
			&& +  \sigma _j^{21}\left( c_{-+}{\bar{c}_{+-}}+c_{++}{\bar{c}_{--}} \right)+\sigma _j^{22}2\operatorname{Re}\left( c_{--}{\bar{c}_{+-}} \right), \nonumber\\
\end{eqnarray}

\begin{eqnarray}	\label{Eq13}
	T_{2j}N && ^{-2} \nonumber\\
		= &&  \sigma _j^{11}2\operatorname{Im}\left( c_{-+}{\bar{c}_{++}} \right)-i\sigma _j^{12}\left( c_{--}{\bar{c}_{++}}-c_{+-}{\bar{c}_{-+}} \right) \nonumber\\
			&& -  i\sigma _j^{21}\left( c_{-+}{\bar{c}_{+-}}-c_{++}{\bar{c}_{--}} \right)+\sigma _j^{22}2\operatorname{Im}\left( c_{--}{\bar{c}_{+-}} \right), \nonumber\\
\end{eqnarray}
\begin{eqnarray}	\label{Eq14}
	T_{3j}N && ^{-2} \nonumber\\
		= && \sigma _j^{11}\left( {{| c_{++} |}^2}-{{| c_{-+} |}^2} \right)+\sigma _j^{12}\left( c_{+-}{\bar{c}_{++}}-c_{--}{\bar{c}_{-+}} \right) \nonumber\\
			&& + \sigma _j^{21}\left( c_{++}{\bar{c}_{+-}}-c_{-+}{\bar{c}_{--}} \right)+\sigma _j^{22}\left( {{| c_{+-} |}^2}-{{| c_{--} |}^2} \right). \nonumber\\
\end{eqnarray}
Using definitions of the Pauli matrices (\ref{Eq08}) once again, from equalities (\ref{Eq12})--(\ref{Eq14}) we obtain the spin correlation tensor for the two-electron system in state (\ref{Eq01}):
\begin{widetext}
\begin{equation}	\label{Eq15}
	T = {N^2}\left[ \begin{matrix}
	2\operatorname{Re}\left( c_{--}{\bar{c}_{++}}+c_{+-}{\bar{c}_{-+}} \right) & 2\operatorname{Im}\left( c_{--}{\bar{c}_{++}}+c_{+-}{\bar{c}_{-+}} \right) & 2\operatorname{Re}\left( c_{-+}{\bar{c}_{++}}-c_{--}{\bar{c}_{+-}} \right) \\
		2\operatorname{Im}\left( c_{--}{\bar{c}_{++}}+c_{-+}{\bar{c}_{+-}} \right) & 2\operatorname{Re}\left( c_{-+}{\bar{c}_{+-}}-c_{--}{\bar{c}_{++}} \right) & 2\operatorname{Im}\left( c_{-+}{\bar{c}_{++}}-c_{--}{\bar{c}_{+-}} \right) \\
		2\operatorname{Re}\left( c_{+-}{\bar{c}_{++}}-c_{--}{\bar{c}_{-+}} \right) & 2\operatorname{Im}\left( c_{+-}{\bar{c}_{++}}-c_{--}{\bar{c}_{-+}} \right) & {{| c_{++} |}^2}-{{| c_{-+} |}^2}-{{| c_{+-} |}^2}+{{| c_{--} |}^2} \\
		\end{matrix} \right].
\end{equation}
\end{widetext}

The tensor (\ref{Eq15}) has the following symmetry that the non-diagonal elements of the tensor $T_{12}$ and $T_{21}$, $T_{13}$ and $T_{31}$, $T_{23}$ and $T_{32}$ are connected with each other:
\begin{equation}	\label{Eq16}
	c_{+-}\mapsto c_{-+},\qquad c_{-+}\mapsto c_{+-}.
\end{equation}

\vfill
\subsection{Tensor in the absence of entanglement}

In the absence of an entanglement, the spin correlation tensor (\ref{Eq04}) is equal to the tensor product of the electron polarization vectors (\textit{polarizations}) \cite{Bib09, Bib11}:
\begin{equation}	\label{Eq17}
	{\tilde{T}}_{ij} := \langle {\sigma}_{i1} \rangle \langle {\sigma}_{j2} \rangle = P_{i1}P_{j2},
\end{equation}
\begin{equation}	\label{Eq18}
	{P_{ia}} := \langle {\sigma}_{ia} \rangle,
\end{equation}
where $P_{ia}$ is a polarization projection of $a$th electron on axis $i$. Let's obtain expressions for polarizations.

According to definition (\ref{Eq07}) and a condition (\ref{Eq02}) for the projections of the electron polarizations in the system in state (\ref{Eq01}) we have
\begin{eqnarray}	\label{Eq19}
	P_{i1}N && ^{-2} \nonumber\\
		= && \sigma _i^{11}\left( {{| c_{++} |}^2}+{{| c_{+-} |}^2} \right)+\sigma _i^{12}\left( {\bar{c}_{++}}c_{-+}+{\bar{c}_{+-}}c_{--} \right) \nonumber\\
			&& + \sigma _i^{21}\left( {\bar{c}_{-+}}c_{++}+{\bar{c}_{--}}c_{+-} \right)+\sigma _i^{22}\left( {{| c_{-+} |}^2}+{{| c_{--} |}^2} \right), \nonumber\\
\end{eqnarray}
\begin{eqnarray}	\label{Eq20}
	P_{j2}N && ^{-2} \nonumber\\
		= &&  \sigma _j^{11}\left( {{| c_{++} |}^2}+{{| c_{-+} |}^2} \right)+\sigma _j^{12}\left( {\bar{c}_{++}}c_{+-}+{\bar{c}_{-+}}c_{--} \right) \nonumber\\
			&& +  \sigma _j^{21}\left( {\bar{c}_{+-}}c_{++}+{\bar{c}_{--}}c_{-+} \right)+\sigma _j^{22}\left( {{| c_{+-} |}^2}+{{| c_{--} |}^2} \right). \nonumber\\
\end{eqnarray}
Taking into account definition of the Pauli matrices (\ref{Eq08}) we can simplify expressions (\ref{Eq19}) and (\ref{Eq20}):
\begin{equation}	\label{Eq21}
	{\bf P}_1 = {N^2}\left[ \begin{matrix}
	2\operatorname{Re}\left( c_{-+}{\bar{c}_{++}}+c_{--}{\bar{c}_{+-}} \right) \\
		2\operatorname{Im}\left( c_{-+}{\bar{c}_{++}}+c_{--}{\bar{c}_{+-}} \right) \\
		{| c_{++} |}^2 + {| c_{+-} |}^2 - {| c_{-+} |}^2 - {| c_{--} |}^2 \\
		\end{matrix} \right],
\end{equation}
\begin{equation}	\label{Eq22}
	{\bf P}_2 = {N^2}\left[ \begin{matrix}
	2\operatorname{Re}\left( c_{+-}{\bar{c}_{++}}+c_{--}{\bar{c}_{-+}} \right) \\
		2\operatorname{Im}\left( c_{+-}{\bar{c}_{++}}+c_{--}{\bar{c}_{-+}} \right) \\
		{| c_{++} |}^2 + {| c_{-+} |}^2 - {| c_{+-} |}^2 - {| c_{--} |}^2 \\
		\end{matrix} \right].
\end{equation}

As well as non-diagonal elements of the spin correlation tensor (\ref{Eq15}), the electron polarizations (\ref{Eq21}) and (\ref{Eq22}) are connected with each other by substituting (\ref{Eq16}).

\section{Tensor measure \\
of quantum entanglement}\label{TensMeas}

As in the absence of entanglement the spin correlation tensor (\ref{Eq04}) is equal to the tensor product (\ref{Eq17}), we set the distance between them as the measure of entanglement in the system. Mathematically the distance between tensors is defined as the norm of their difference:
\begin{equation}	\label{Eq23}
	E := ||T-\tilde{T}||.
\end{equation}
As the norm we choose the scaled Euclidean norm:
\begin{equation}	\label{Eq24}
	\forall A\in \mathbb{R}^{3\times 3}\quad ||A|| = \sqrt{\operatorname{tr}\left( A{A^T} \right)/3} = \sqrt{{A_{ij}}{A_{ij}}/3}.
\end{equation}
Here and everywhere below $\operatorname{tr}$ is a trace of the real $3\times 3$ matrices.

The measure of the quantum entanglement is valid when it has the following properties \cite{Bib04, Bib05}:

1.	\textit{nonnegativity, discriminance, normalization};

2.	\textit{invariance under local unitary operations (LU)};

3.	\textit{non-growth under measurements};

4.	\textit{non-growth under local operations and classical communication (LOCC)}.

Let's prove these properties for the measure (\ref{Eq23}). For this purpose we will use a definition of mean values in terms of a density matrix (\ref{Eq06}).

\subsection{Nonnegativity, discriminance, normalization}

\textit{Proposition 1}. The measure (\ref{Eq23}) is nonnegative (\textit{nonnegativity}):
\begin{equation}	\label{Eq25}
	\forall \rho \quad E(\rho)\ge 0.
\end{equation}

\textit{Proof}. Property (\ref{Eq25}) is carried out for the measure (\ref{Eq23}) by definition of the norm.~\hfill$\blacksquare$

\textit{Proposition 2}. The criterion for separability of states is that the measure (\ref{Eq23}) is equal to zero (\textit{discriminance}):
\begin{equation}	\label{Eq26}
	\rho \text{ is separable}\quad \Leftrightarrow \quad E(\rho) = 0.
\end{equation}

\textit{Proof}. According to the proposition 1a in the work \cite{Bib12}, taking into account that for a system of electrons the Bloch vector coincides with polarization by definition \cite{Bib07}, we have
\begin{equation}	\label{Eq27}
	\rho \text{ is separable}\quad \Leftrightarrow \quad T = {\bf P}_1 \otimes {\bf P}_2.
\end{equation}
Property (\ref{Eq26}) for the measure (\ref{Eq23}) follows from the statement (\ref{Eq27}) and definition (\ref{Eq17}).~\hfill$\blacksquare$

\textit{Proposition 3}. The measure (\ref{Eq23}) for maximally entangled states is equal to one (\textit{normalization}):
\begin{equation}	\label{Eq28}
	\rho \text{ is maximally entangled}\quad \Rightarrow \quad E(\rho) = 1.
\end{equation}

\textit{Proof}. For two-particle system maximally entangled states are Bell states \cite{Bib05}. In terms of function (\ref{Eq01}) one can write them as
\begin{equation}	\label{Eq29}
	\begin{aligned}
		N = \tfrac{1}{\sqrt{2}}, \ c_{++} = 0, \ c_{+-} = +1, & \ c_{-+} = -1, \ c_{--} = 0; \\
		N = \tfrac{1}{\sqrt{2}}, \ c_{++} = 0, \ c_{+-} = +1, & \ c_{-+} = +1, \ c_{--} = 0; \\
		N = \tfrac{1}{\sqrt{2}}, \ c_{++} = +1, \ c_{+-} = 0, & \ c_{-+} = 0, \ c_{--} = -1; \\
		N = \tfrac{1}{\sqrt{2}}, \ c_{++} = +1, \ c_{+-} = 0, & \ c_{-+} = 0, \ c_{--} = +1; \\
	\end{aligned}
\end{equation}
for a singlet and three triplets respectively. From equalities (\ref{Eq29}), expressions (\ref{Eq21}), (\ref{Eq22}) and (\ref{Eq15}) follows that in all four conditions polarizations are equal to zero, and spin correlation tensors are diagonal matrices which elements are equal to $\pm 1$. Then taking into account definition (\ref{Eq17}) for the measure (\ref{Eq23}) the property (\ref{Eq28}) is carried out.~\hfill$\blacksquare$

\subsection{LU invariance}

\textit{Proposition 4}. The measure (\ref{Eq23}) is invariant under local unitary operations:
\begin{equation}	\label{Eq30}
	E(\rho) = E({\rho}'),
\end{equation}
\begin{equation}	\label{Eq31}
	{\rho}' := (U_1^{\dagger}\otimes U_2^{\dagger})\rho ({U_1}\otimes {U_2}).
\end{equation}
Here $U_a$ is the unitary operator acting on $a$th particle:
\begin{equation}	\label{Eq32}
	{U_a}U_a^{\dagger} = U_a^{\dagger}{U_a} = {I_a},
\end{equation}
where $I_a$ is the unity operator acting on $a$th particle.

\textit{Proof}. After transformation, the measure (\ref{Eq23}) takes the form:
\begin{equation}	\label{Eq33}
	E({\rho}') = ||{T}'-{\tilde{T}}'||,
\end{equation}
where tensors (\ref{Eq04}) and (\ref{Eq17}) can be written in terms of a density matrix ${\rho}'$ (\ref{Eq06}). Then taking into account expression (\ref{Eq31}) and properties of a trace we have
\begin{eqnarray*}
	T'_{ij} &=& \operatorname{Tr}\left( {\rho}'{\sigma}_{i1}{\sigma}_{j2} \right) \\
		&=& \operatorname{Tr}\left[ \rho ({U_1}\otimes {U_2}){\sigma}_{i1}{\sigma}_{j2}(U_1^{\dagger}\otimes U_2^{\dagger}) \right]\quad \Rightarrow
\end{eqnarray*}
\begin{equation}	\label{Eq34}
	T'_{ij} = \operatorname{Tr}\left[ \rho ({U_1}{\sigma}_{i1}U_1^{\dagger})({U_2}{\sigma}_{j2}U_2^{\dagger}) \right].
\end{equation}

The Pauli matrices form a basis in the space of Hermitian $2\times 2$ matrices with a zero trace. At the same time taking into account unitarity (\ref{Eq32}) $\operatorname{Tr}\left( {U_a}{\sigma}_{ia}U_a^{\dagger} \right)$ $= \operatorname{Tr}\left( U_a^{\dagger}{U_a}{\sigma}_{ia} \right) = \operatorname{Tr}\left( {\sigma}_{ia} \right) = 0$. Then in the expression (\ref{Eq34}), the quantity ${U_a}{\sigma}_{ia}U_a^{\dagger}$ can be expanded on the basis:
\begin{equation}	\label{Eq35}
	{U_a}{\sigma}_{ia}U_a^{\dagger} =: {Q_{ija}}{{\sigma}_{ja}},
\end{equation}
where $Q_a$ is orthogonal $3\times 3$ matrix (see Appendix \ref{PropQD}):
\begin{equation}	\label{Eq36}
	{Q_a}Q_a^T = Q_a^T{Q_a} = I.
\end{equation}
At the same time the matrix $Q_a$ can be taken out from under the trace $\operatorname{Tr}$.

From expression (\ref{Eq34}) and the expansion (\ref{Eq35}) we have:
\begin{eqnarray*}
	T'_{ij} &=& \operatorname{Tr}\left[ \rho ({Q_{ik1}}{{\sigma}_{k1}})({Q_{jl2}}{{\sigma}_{l2}}) \right] \\
		&=& {Q_{ik1}}{Q_{jl2}}\operatorname{Tr}\left( \rho {{\sigma}_{k1}}{{\sigma}_{l2}} \right) = {Q_{ik1}}{Q_{jl2}}{T_{kl}}\quad \Rightarrow
\end{eqnarray*}
\begin{equation}	\label{Eq37}
	T' = {Q_1}{{\times}_1}{Q_2}{{\times}_2}T,
\end{equation}
where symbol ${\times}_a$ means action of a matrix (on the left) on $a$th index of a tensor (on the right). Taking into account the expression (\ref{Eq31}), properties of a trace, the unitarity (\ref{Eq32}) and the expansion (\ref{Eq35}) we obtain:
\begin{eqnarray*}
	P'_{ia} &=& \operatorname{Tr}\left( {\rho}'{\sigma}_{ia} \right) = \operatorname{Tr}\left[ \rho ({U_1}\otimes {U_2}){\sigma}_{ia}(U_1^{\dagger}\otimes U_2^{\dagger}) \right] \\
		&=& \operatorname{Tr}\left[ \rho ({U_a}{\sigma}_{ia}U_a^{\dagger}) \right] = {Q_{ika}}\operatorname{Tr}\left( \rho {{\sigma}_{ka}} \right) = {Q_{ika}}{P_{ka}}.
\end{eqnarray*}
Then for the transformed tensor product (\ref{Eq17}) we have
\begin{eqnarray*}
	{\tilde{T}'}_{ij} = {Q_{ik1}}{P_{k1}}{Q_{jl2}}{P_{l2}} = {Q_{ik1}}{Q_{jl2}}{{\tilde{T}}_{kl}}\quad \Rightarrow
\end{eqnarray*}
\begin{equation}	\label{Eq38}
	\tilde{T'} = {Q_1}{{\times}_1}{Q_2}{{\times}_2}\tilde{T}.
\end{equation}
Taking into account expressions (\ref{Eq37}) and (\ref{Eq38}) the transformed measure (\ref{Eq33}) takes the form:
\begin{equation}	\label{Eq39}
	E({\rho}') = ||{Q_1}{{\times}_1}{Q_2}{{\times}_2}(T-\tilde{T})||.
\end{equation}

From expressions (\ref{Eq39}) and (\ref{Eq24}), properties of a trace and the orthogonality (\ref{Eq36}) we have
\begin{eqnarray*}
	3 && {{\left[ E({\rho}') \right]}^2} \\
		&&= \operatorname{tr}\left\{ [{Q_1}{{\times}_1}{Q_2}{{\times}_2}(T-\tilde{T})]{[{Q_1}{{\times}_1}{Q_2}{{\times}_2}(T-\tilde{T})]}^T \right\} \\
		&&= \operatorname{tr}\left\{ {Q_1}[{Q_2}{{\times}_2}(T-\tilde{T})]{[{Q_2}{{\times}_2}(T-\tilde{T})]}^T Q_1^T \right\} \\
		&&= \operatorname{tr}\left\{ [{Q_2}{{\times}_2}(T-\tilde{T})]{[{Q_2}{{\times}_2}(T-\tilde{T})]}^T \right\} \\
		&&= \operatorname{tr}\left\{ {{[{Q_2}{(T-\tilde{T})^T}]}^T}[{Q_2}{(T-\tilde{T})^T}] \right\} \\
		&&= \operatorname{tr}\left[ (T-\tilde{T})Q_2^T{Q_2}{(T-\tilde{T})^T} \right] \\
		&&= \operatorname{tr}\left[ (T-\tilde{T})(T-\tilde{T})^T \right] = 3||T-\tilde{T}||^2,
\end{eqnarray*}
from where the property (\ref{Eq30}) of the measure (\ref{Eq23}) follows.~\hfill$\blacksquare$

\subsection{Non-growth under measurements}

\textit{Proposition 5}. The measure (\ref{Eq23}) does not increase under measurements:
\begin{equation}	\label{Eq40}
	E(\rho)\ge E({\rho}'),
\end{equation}
where $\rho $ is a density matrix of the original two-particle pure state, ${\rho}'$ is a density matrix of the resulting two-particle mixed state. Without losing generality, we assume that the local measurements are \textit{positive operator value measures} (\textit{POVMs}) \cite{Bib05}. The local POVMs acting on a two-particle state generally have an appearance:
\begin{equation}	\label{Eq41}
	{\rho}' := \sum\nolimits_{mn}{({L_{m1}}\otimes {L_{n2}})\rho (L_{m1}^{\dagger}\otimes L_{n2}^{\dagger})},
\end{equation}
where ${\{{L_{na}}\}}_n$ are linear, positive, keeping a trace operators having properties:
\begin{equation}	\label{Eq42}
	\sum\nolimits_n{{L_{na}}L_{na}^{\dagger}} = {I_a},\quad [{L_{na}},L_{na}^{\dagger}] = 0.
\end{equation}

\textit{Proof}. After transformation, the measure (\ref{Eq23}) takes the form:
\begin{equation}	\label{Eq43}
	E({\rho}') = ||{T}'-{\tilde{T}}'||,
\end{equation}
where tensors (\ref{Eq04}) and (\ref{Eq17}) can be written in terms of a density matrix ${\rho}'$ (\ref{Eq06}). Then taking into account expression (\ref{Eq41}) and properties of a trace we have
\begin{eqnarray*}
	T'_{ij} &=& \operatorname{Tr}\left( {\rho}'{\sigma}_{i1}{\sigma}_{j2} \right) \\
		&=& \sum\nolimits_{mn}{\operatorname{Tr}\left[ \rho (L_{m1}^{\dagger}\otimes L_{n2}^{\dagger}){\sigma}_{i1}{\sigma}_{j2}({L_{m1}}\otimes {L_{n2}}) \right]}\,\Rightarrow
\end{eqnarray*}
\begin{equation}	\label{Eq44}
	T'_{ij} = \operatorname{Tr}\left[ \rho \left( \sum\nolimits_m{L_{m1}^{\dagger}{\sigma}_{i1}{L_{m1}}} \right)\left( \sum\nolimits_n{L_{n2}^{\dagger}{\sigma}_{j2}{L_{n2}}} \right) \right].
\end{equation}

The Pauli matrices form a basis in the space of Hermitian $2\times 2$ matrices with a zero trace. At the same time taking into account properties (\ref{Eq42}) $\operatorname{Tr}\left( \sum\nolimits_n{L_{na}^{\dagger}{\sigma}_{ia}{L_{na}}} \right) = \sum\nolimits_n{\operatorname{Tr}\left( {L_{na}}L_{na}^{\dagger}{\sigma}_{ia} \right)} = \operatorname{Tr}\left( {\sigma}_{ia} \right)$ $= 0$. Then in the expression (\ref{Eq44}), the quantity $\sum\nolimits_n{L_{na}^{\dagger}{\sigma}_{ia}{L_{na}}}$ can be expanded on the basis:
\begin{equation}	\label{Eq45}
	\sum\nolimits_n{L_{na}^{\dagger}{\sigma}_{ia}{L_{na}}} = :{D_{ija}}{{\sigma}_{ja}},
\end{equation}
where $D_a$ is real contractive $3\times 3$ matrix (see Appendix~\ref{PropQD}):
\begin{equation}	\label{Eq46}
	{D_a}D_a^T = D_a^T{D_a}\le I.
\end{equation}
At the same time the matrix $D_a$ can be taken out from under the trace $\operatorname{Tr}$.

From expression (\ref{Eq44}) and the expansion (\ref{Eq45}) we have
\begin{eqnarray*}
	T'_{ij} &=& \operatorname{Tr}\left[ \rho ({D_{ik1}}{{\sigma}_{k1}})({D_{jl2}}{{\sigma}_{l2}}) \right] \\
		&=& {D_{ik1}}{D_{jl2}}\operatorname{Tr}\left( \rho {{\sigma}_{k1}}{{\sigma}_{l2}} \right) = {D_{ik1}}{D_{jl2}}{T_{kl}}\quad \Rightarrow
\end{eqnarray*}
\begin{equation}	\label{Eq47}
	T' = {D_1}{{\times}_1}{D_2}{{\times}_2}T.
\end{equation}
Taking into account the expression (\ref{Eq41}), properties of a trace, properties (\ref{Eq42}) and the expansion (\ref{Eq45}) we obtain:
\begin{eqnarray*}
	P'_{i1} &=& \operatorname{Tr}\left( {\rho}'{\sigma}_{i1} \right) \\
		&=& \sum\nolimits_{mn}{\operatorname{Tr}\left[ \rho (L_{m1}^{\dagger}\otimes L_{n2}^{\dagger}){\sigma}_{i1}({L_{m1}}\otimes {L_{n2}}) \right]} \\
		&=& \sum\nolimits_{mn}{\operatorname{Tr}\left[ \rho (L_{m1}^{\dagger}{\sigma}_{i1}{L_{m1}})(L_{n2}^{\dagger}{L_{n2}}) \right]} \\
		&=& \operatorname{Tr}\left[ \rho \left( \sum\nolimits_m{L_{m1}^{\dagger}{\sigma}_{i1}{L_{m1}}} \right) \right] = {D_{ik1}}\operatorname{Tr}\left( \rho {{\sigma}_{k1}} \right) \\
		&=& {D_{ik1}}{P_{k1}}.
\end{eqnarray*}
One also can show that ${P'_{j2}} = {D_{jl2}}{P_{l2}}$. Then for the transformed tensor product (\ref{Eq17}) we have
\begin{eqnarray*}
	{\tilde{T}'}_{ij} = {D_{ik1}}{P_{k1}}{D_{jl2}}{P_{l2}} = {D_{ik1}}{D_{jl2}}{{\tilde{T}}_{kl}}\quad \Rightarrow
\end{eqnarray*}
\begin{equation}	\label{Eq48}
	{\tilde{T}}' = {D_1}{{\times}_1}{D_2}{{\times}_2}\tilde{T}.
\end{equation}
Taking into account expressions (\ref{Eq47}) and (\ref{Eq48}) the transformed measure (\ref{Eq43}) takes the form:
\begin{equation}	\label{Eq49}
	E({\rho}') = ||{D_1}{{\times}_1}{D_2}{{\times}_2}(T-\tilde{T})||.
\end{equation}

From expressions (\ref{Eq49}) and (\ref{Eq24}), properties of a trace and the property (\ref{Eq46}) we have
\begin{eqnarray*}
	3 && {{\left[ E({\rho}') \right]}^2} \\
		&&= \operatorname{tr}\left\{ [{D_1}{{\times}_1}{D_2}{{\times}_2}(T-\tilde{T})]{{[{D_1}{{\times}_1}{D_2}{{\times}_2}(T-\tilde{T})]}^T} \right\} \\
		&&= \operatorname{tr}\left\{ {D_1}[{D_2}{{\times}_2}(T-\tilde{T})]{{[{D_2}{{\times}_2}(T-\tilde{T})]}^T}D_1^T \right\} \\
		&&\le \operatorname{tr}\left\{ [{D_2}{{\times}_2}(T-\tilde{T})]{{[{D_2}{{\times}_2}(T-\tilde{T})]}^T} \right\} \\
		&&= \operatorname{tr}\left\{ {[{D_2}{(T-\tilde{T})^T}]}^T[{D_2}(T-\tilde{T})^T] \right\} \\
		&&= \operatorname{tr}\left[ (T-\tilde{T})D_2^T{D_2}(T-\tilde{T})^T \right] \\
		&&\le \operatorname{tr}\left[ (T-\tilde{T})(T-\tilde{T})^T \right] = 3||T-\tilde{T}||^2,
\end{eqnarray*}
from where the property (\ref{Eq40}) of the measure (\ref{Eq23}) follows.~\hfill$\blacksquare$

\subsection{Non-growth under LOCC}

\textit{Proposition 6}. The measure (\ref{Eq23}) does not increase under local operations and classical communication ${\Phi}_{\text{LOCC}}$:
\begin{equation}	\label{Eq50}
	E(\rho)\ge E \textbf{(}{{\Phi}_{\text{LOCC}}}(\rho) \textbf{)}.
\end{equation}

\textit{Proof}. LOCC can be decomposed into four basic kinds of operations \cite{Bib13}.

I. \textit{Appending an ancillary system not entangled to the state of the original system}. It is obvious that appending cannot change the tensor (\ref{Eq04}) and polarizations (\ref{Eq18}). Therefore the measure (\ref{Eq23}) is invariant under appending.

II. \textit{Performing a unitary transformation}. The measure (\ref{Eq23}) is invariant under the unitary transformations (\ref{Eq30}).

III. \textit{Performing measurements}. The measure (\ref{Eq23}) does not increase under the measurements (\ref{Eq40}).

IV. \textit{Throwing away (tracing out) part of the system}. It is obvious that after this operation in two-particle system the entanglement is equal to zero.

As for the measure (\ref{Eq23}) all 4 requirements are fulfilled, for it property (\ref{Eq50}) is true.~\hfill$\blacksquare$

Thus, according to the properties proved in this section, the measure of quantum entanglement (\ref{Eq23}) is valid.

\section{Tensor measure \\
of quantum entanglement \\
in a scattering problem}

\subsection{Electron-electron scattering problem}

Let's calculate measure of quantum entanglement (\ref{Eq23}) in a problem of Coulomb electron-electron scattering \cite{Bib10}. In this case, in expression (\ref{Eq01}) one should set
\begin{equation}	\label{Eq51}
	\begin{aligned}
		c_{++} &= \cos \left( | \Omega |/2 \right){{\psi}_a}, \\
		c_{+-} &= \tfrac{1}{2}{e^{i\varphi}}\sin \left( | \Omega |/2 \right)\left( {{\psi}_s}+{{\psi}_a} \right), \\
		c_{-+} &= \tfrac{1}{2}{e^{i\varphi}}\sin \left( | \Omega |/2 \right)\left( {{\psi}_a}-{{\psi}_s} \right), \\
		c_{--} &= 0, \\
	\end{aligned}
\end{equation}
\begin{equation}	\label{Eq52}
	{{\psi}_s}(\theta) = f(\theta)+f(\pi -\theta),\quad {{\psi}_a}(\theta) = f(\theta)-f(\pi -\theta),
\end{equation}
\begin{equation}	\label{Eq53}
	f(\theta)\sim \csc {{(\theta /2)}^2}\exp [-i\alpha \ln (1-\cos \theta)].
\end{equation}
In expressions (\ref{Eq51})--(\ref{Eq53}) $\Omega $ and $\varphi $ are polar (relative to the axis $z$) and azimuthal (relative to the axis $x$) rotation angles of the 2nd electron polarization before scattering (polarization of the 1st electron is oriented in the $z$-direction), ${\psi}_s$ is symmetric wave function, ${\psi}_a$ is anti-symmetric wave function, $f$ is the scattering amplitude in the centre of mass of the interacting electrons, $\theta $ is scattering angle in the centre of mass frame, $\alpha = 1/{{\upsilon}_{\text{rel}}}$ is the dimensionless factor, ${\upsilon}_{\text{rel}}$ is the relative electron velocity in atomic units.

\subsection{$\varphi $ independence of entanglement measure}

\textit{Proposition 7}. The measure (\ref{Eq23}) in the problem (\ref{Eq01}), (\ref{Eq51})--(\ref{Eq53}) does not depend on azimuthal angle:
\begin{equation}	\label{Eq54}
	E = \text{const}(\varphi).
\end{equation}

\textit{Proof}. Let's segregate the sum in the measure (\ref{Eq23}) into three blocks:
\begin{equation}	\label{Eq55}
	3E^2 = T_{ij}T_{ij}-2T_{ij}{\tilde{T}}_{ij}+{\tilde{T}}_{ij}{\tilde{T}}_{ij}.
\end{equation}
At the same time everywhere below we will consider that according to expressions (\ref{Eq03}) and (\ref{Eq51}) $N = \text{const}(\varphi)$.

Let's consider the block $T_{ij}T_{ij}$. According to expressions (\ref{Eq15}) and (\ref{Eq51}) elements $T{{_{xx}^2}^{}}$, $T_{xy}^2$, $T_{yx}^2$, $T_{yy}^2$ and $T_{zz}^2$ do not depend on $\varphi $, and also one can see that
\begin{eqnarray*}
	 T_{xz}^2+T_{yz}^2 &=& {{| {N^2}2c_{-+}{\bar{c}_{++}} |}^2} = \text{const}(\varphi), \\
	 T_{zx}^2+T_{zy}^2 &=& {{| {N^2}2c_{+-}{\bar{c}_{++}} |}^2} = \text{const}(\varphi).
\end{eqnarray*}
Consequently, the block $T_{ij}T_{ij}$ does not depend on $\varphi $:
\begin{equation}	\label{Eq56}
	T_{ij}T_{ij} = \text{const}(\varphi).
\end{equation}

Let's consider the block $T_{ij}{\tilde{T}}_{ij}$. From expressions (\ref{Eq15}), (\ref{Eq21}) and (\ref{Eq22}) taking into account equalities (\ref{Eq51}) follows that ${T_{zz}}{{\tilde{T}}_{zz}} = {T_{zz}}{P_{z1}}{P_{z2}} = \text{const}(\varphi)$. Also from them one can see that ${T_{xz}} = {P_{x1}}$, ${T_{yz}} = {P_{y1}}$, ${T_{zx}} = {P_{x2}}$, ${T_{zy}} = {P_{y2}}$. Then
\begin{eqnarray*}
	T_{xz}{\tilde{T}}_{xz} &+& T_{yz}{\tilde{T}}_{yz} = (P_{x1}^2 + P_{y1}^2){P_{z2}} \\
		&=& {{| {N^2}2c_{-+}{\bar{c}_{++}} |}^2}P_{z2} = \text{const}(\varphi), \\
	T_{zx}{\tilde{T}}_{zx} &+& T_{zy}{\tilde{T}}_{zy} = {P_{z1}}(P_{x2}^2 + P_{y2}^2) \\
		&=& P_{z1}{{| {N^2}2c_{+-}{\bar{c}_{++}} |}^2} = \text{const}(\varphi).
\end{eqnarray*}
Taking into account equalities (\ref{Eq15}), (\ref{Eq17}), (\ref{Eq21}) and (\ref{Eq22}) we have:
\begin{eqnarray*}
	T_{xx} && {\tilde{T}}_{xx} + T_{yx}{\tilde{T}}_{yx} \\
		&&= {N^4}2(c_{-+}c_{+-}{\bar{c}_{-+}}{\bar{c}_{++}}+c_{-+}c_{++}{\bar{c}_{-+}}{\bar{c}_{+-}}){P_{x2}}, \\
	T_{xy} && {\tilde{T}}_{xy} + T_{yy}{\tilde{T}}_{yy} \\
		&&= -i{N^4}2(c_{-+}c_{+-}{\bar{c}_{-+}}{\bar{c}_{++}}-c_{-+}c_{++}{\bar{c}_{-+}}{\bar{c}_{+-}}){P_{y2}}.
\end{eqnarray*}
Substituting in these expressions $P_{x2}$ and $P_{y2}$ (\ref{Eq22}), taking into account equalities (\ref{Eq51}) we see that
\begin{eqnarray*}
	T_{xx}{\tilde{T}}_{xx} &+& T_{yx}{\tilde{T}}_{yx} + T_{xy}{\tilde{T}}_{xy} + T_{yy}{\tilde{T}}_{yy} \\
		&=& {N^6}8{{| c_{-+}c_{+-}c_{++} |}^2} = \text{const}(\varphi).
\end{eqnarray*}
Consequently, the block $T_{ij}{\tilde{T}}_{ij}$ does not depend on $\varphi $:
\begin{equation}	\label{Eq57}
	T_{ij}{\tilde{T}}_{ij} = \text{const}(\varphi).
\end{equation}

Let's consider the block ${\tilde{T}}_{ij}{\tilde{T}}_{ij}$. From expressions (\ref{Eq17}), (\ref{Eq21}), (\ref{Eq22}) and (\ref{Eq51}) we have
\begin{eqnarray*}
	\tilde{T}_{xx}^2 &+& \tilde{T}_{yx}^2 + \tilde{T}_{xy}^2 + \tilde{T}_{yy}^2 = (P_{x1}^2 + P_{y1}^2)(P_{x2}^2 + P_{y2}^2) \\
		&=& {{| {N^2}2c_{-+}{\bar{c}_{++}} |}^2}{{| {N^2}2c_{+-}{\bar{c}_{++}} |}^2} = \text{const}(\varphi), \\
	\tilde{T}_{xz}^2 &+& \tilde{T}_{yz}^2 = (P_{x1}^2 + P_{y1}^2)P_{z2}^2 \\
		&=& {{| {N^2}2c_{-+}{\bar{c}_{++}} |}^2}P_{z2}^2 = \text{const}(\varphi), \\
	\tilde{T}_{zx}^2 &+& \tilde{T}_{zy}^2 = P_{z1}^2(P_{x2}^2 + P_{y2}^2) \\
		&=& P_{z1}^2{{| {N^2}2c_{+-}{\bar{c}_{++}} |}^2} = \text{const}(\varphi), \\
	\tilde{T}_{zz}^2 &=& P{{_{z1}^2}_{}}P_{z2}^2 = \text{const}(\varphi).
\end{eqnarray*}
Consequently, the block ${\tilde{T}}_{ij}{\tilde{T}}_{ij}$ does not depend on $\varphi $:
\begin{equation}	\label{Eq58}
	{\tilde{T}}_{ij}{\tilde{T}}_{ij} = \text{const}(\varphi).
\end{equation}
Thus, property (\ref{Eq54}) for measure (\ref{Eq23}) follows from the equalities (\ref{Eq55})--(\ref{Eq58}).~\hfill$\blacksquare$

\subsection{Numerical calculation}

\begin{figure}[htb]\center
	\includegraphics{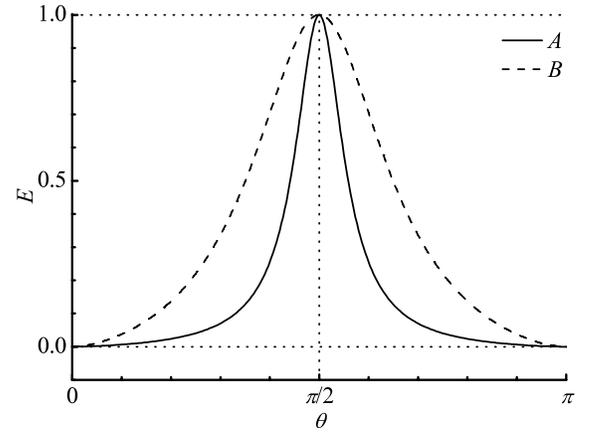}
	\caption{The quantum entanglement measure (\ref{Eq23}) of two scattered electrons in state (\ref{Eq01}), (\ref{Eq51})--(\ref{Eq53}) at ${{\upsilon}_{\text{rel}}} = 1.5$, $\varphi = 0$ (does not depend on $\varphi $). Line $A$: $\Omega = \pi /4$; line $B$: $\Omega = 3\pi /4$.}\label{Fig01}
\end{figure}

\begin{figure}[htb]\center
	\includegraphics{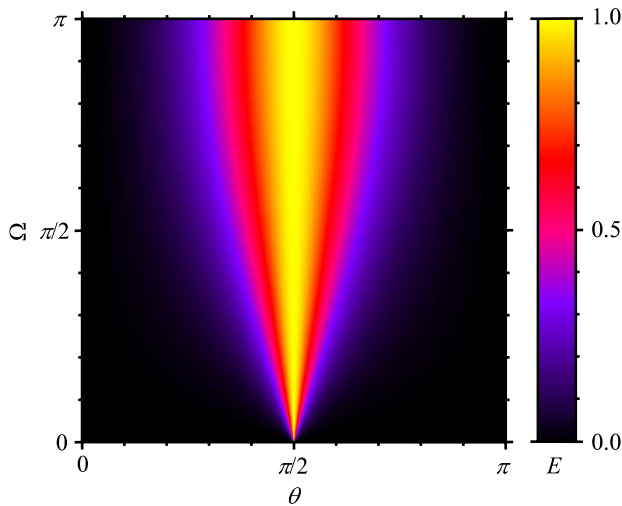}
	\caption{The quantum entanglement measure of two scattered electrons in state (\ref{Eq01}), (\ref{Eq51})--(\ref{Eq53}) at ${{\upsilon}_{\text{rel}}} = 1.5$, $\varphi = 0$ (does not depend on $\varphi $).}\label{Fig02}
\end{figure}

Let's plot the graph of quantum entanglement measure (\ref{Eq23}) taking into account equalities (\ref{Eq24}), (\ref{Eq15}), (\ref{Eq17}), (\ref{Eq21}), (\ref{Eq22}), (\ref{Eq51})--(\ref{Eq53}) (Figs.\ref{Fig01} and \ref{Fig02}). Calculations confirmed that it does not depend on the azimuthal angle $\varphi $. Therefore, without losing the generality, we set $\varphi = 0$ in the following figures. From them we see that the entanglement reaches a maximum at $\theta = \pi /2$, where it is equal to one. The entanglement is equal to zero at $\theta = 0,\pi $. One can observe a peak broadening about a point $\theta = \pi /2$ with increase of angle $\Omega $. These results completely conform and supplement the ones obtained in article \cite{Bib10} where the entropy of one of the electrons of the correlated electron pair was the measure of the entanglement.

\section{Tensor measure \\
of quantum entanglement \\
in experiment}\label{TensExp}

The measure of quantum entanglement (\ref{Eq23}) in the two-electron system can be found from the experiment on electron-electron scattering. For this purpose, it is necessary to measure the spin correlation tensor and the electron polarizations. In this section they act as phenomenological quantities. One can measure them even for a system, which has no microscopic model, and give the quantitative assessment to a quantum entanglement in it.

The spin correlation tensor (\ref{Eq04}) and the electron polarizations (\ref{Eq18}) can be found on the basis of the dimensionless projections of electron spins to coordinate axes measured in experiment. After each act of scattering both electrons are detected separately by two analyzers. Analyzers register the spin projections after $m$th act of scattering $\sigma _{i1}^{(m)}$ and $\sigma _{j2}^{(m)}$ on the axes chosen for them $i$ and $j$ respectively. Spin projections take on values $\pm 1$ that corresponds to eigenvalues of operators (\ref{Eq07}).

In experiment, we measure projections of electron spins at various combinations of the axes chosen for analyzers. Data of the experiment are registered in the form of $3\times 3$ matrix $\mathbb{T} = {{[{{\mathbb{T}}_{ij}}]}_{ij}}$. Each element of the matrix is the table of two columns and $M_{ij}$ rows of data:
\begin{eqnarray}	\label{Eq59}
	{\mathbb{T}}_{ij} := [ && (\sigma _{i1}^{(m)},\sigma _{j2}^{(m)}): \nonumber\\
		&& \sigma _{i1}^{(m)},\sigma _{j2}^{(m)}\in \{+1,-1\},\ m = 1,...,{M_{ij}}].
\end{eqnarray}
The table (\ref{Eq59}) corresponds to the set of measurements of electron spins projections on axes $i$ and $j$. It allows to find the corresponding element of spin correlation tensor (\ref{Eq04}) on the basis of the products $\sigma _{i1}^{(m)}\sigma _{j2}^{(m)}$. As well as spin projections, their products take on values $\pm 1$ that corresponds to the eigenvalues of the operator ${\sigma}_{i1}{\sigma}_{j2}$ (see Appendix \ref{EigVal}).

When calculating projections of electron polarizations (\ref{Eq18}) we use data (\ref{Eq59}). Minimum errors of these quantities can be expected if all available data for the corresponding spin projections are used:
\begin{equation}	\label{Eq60}
	\begin{aligned}
		{{\mathbb{S}}_{i1}} := [{{\bigcup}_j}{\mathbb{T}}_{ij}](,1), \ \operatorname{card}({{\mathbb{S}}_{i1}}) &= N_{i1} := {{\Sigma}_j}M_{ij}, \\
		{{\mathbb{S}}_{j2}} := [{{\bigcup}_i}{\mathbb{T}}_{ij}](,2), \ \operatorname{card}({{\mathbb{S}}_{j2}}) &= N_{j2} := {{\Sigma}_i}M_{ij}. \\
	\end{aligned}
\end{equation}
The array ${\mathbb{S}}_{i1}$ is the first column of the table, which is obtained by combining all elements of the matrix $\mathbb{T}$ in the row $i$. The array ${\mathbb{S}}_{j2}$ is the second column of the table, which is obtained by combining all elements of the matrix $\mathbb{T}$ in the column $j$.

At rather large numbers of measurements ${\{{M_{ij}}\}}_{ij}$ from tables (\ref{Eq59}) and arrays (\ref{Eq60}) we have for spin correlation tensor (\ref{Eq04}) and projections of electron polarizations (\ref{Eq18})
\begin{equation}	\label{Eq61}
	\langle {\sigma}_{i1}{\sigma}_{j2} \rangle = \frac{1}{{M_{ij}}}\sum\nolimits_m{(1\le m\le {M_{ij}})\sigma _{i1}^{(m)}\sigma _{j2}^{(m)}},
\end{equation}
\begin{equation}	\label{Eq62}
	\langle {\sigma}_{ia} \rangle = \frac{1}{{N_{ia}}}\sum\nolimits_n{(1\le n\le {N_{ia}})\sigma _{ia}^{(n)}},
\end{equation}
respectively. In expressions (\ref{Eq61}) and (\ref{Eq62}) Iverson notation is used \cite{Bib14}: brackets with the statement are equal to one if it is true, and are equal to zero if it is false. Taking into account definition (\ref{Eq17}) these expressions allow finding the measure (\ref{Eq23}). Thus, the quantum entanglement in the two-electron system can be measured in an experiment on electron-electron scattering.

\section{Conclusion}

In this work, we considered the problem of correct measurement of a quantum entanglement in the two-body electron-electron scattering. An expression is derived for a spin correlation tensor in case of a pure two-electron state. On its basis, we proposed geometrical measure of a quantum entanglement in a system of two particles. It is the distance between two forms of this tensor: in the entangled and separable cases, that makes this measure obvious. As distance between tensors, we used the scaled Euclidean norm of their difference. It enabled the proof of the properties of a measure confirming its validity: nonnegativity, discriminance, normalization, non-growth under local operations and classical communication.

We calculated the measure of quantum entanglement suggested in this paper for a problem of Coulomb electron-electron scattering. We revealed numerically and proved analytically that it does not depend on the azimuthal rotation angle of the second electron spin relative to the first electron spin before scattering. We also suggested a procedure of measurement of a spin correlation tensor in the electron-electron scattering. It allows finding a measure of a quantum entanglement in the experiment even in the absence of a microscopic model of the studied system.

Thus, the main positive characteristics of the tensor measure of a quantum entanglement proposed in this work are:

$\bullet$	\textit{obviousness};

$\bullet$	\textit{validity};

$\bullet$	\textit{measurability in an experiment}.

Prospects of further development of this approach are bound to its generalization on a case of the mixed and multi-electron states. The suggested experimental procedure sets a direction for the correct measurement of quantum entanglement in electron-electron scattering.

\begin{acknowledgments}
The authors acknowledge Saint-Petersburg State University for a research grant 11.38.187.2014 and the University of Western Australia for support.
\end{acknowledgments}

\appendix

\section{properties of matrices $Q_a$ and $D_a$}\label{PropQD}

In this appendix, we will prove properties of matrices $Q_a$ (\ref{Eq35}) and $D_a$ (\ref{Eq45}) which we used in the section \ref{TensMeas} at the proof of properties of the measure (\ref{Eq23}) (propositions 4 and 5). We will carry out proofs on the basis of expression for products of the Pauli matrices:
\begin{equation}	\label{Eq63}
	{\sigma}_{ia}{\sigma}_{ja} = i{\epsilon}_{ijk}{\sigma}_{ka} + I_{ij}I_a,
\end{equation}
where ${\epsilon}_{ijk}$ is Levi-Civita symbol. Having picked up a trace from expression (\ref{Eq63}) on pairs of one-electron states ${{| \pm \rangle}_1}{{| \pm \rangle}_2}$ and ${{| \pm \rangle}_1}{{| \mp \rangle}_2}$, we have
\begin{equation}	\label{Eq64}
	\operatorname{Tr}\left( {\sigma}_{ia}{\sigma}_{ja} \right) = {I_{ij}}\operatorname{Tr}\left( I_a \right).
\end{equation}
Everywhere below we use expression (\ref{Eq64}) and properties of the trace. For brevity we also use references to formulas without their names, symbols of an implication $\Rightarrow $ and biconditional $\Leftrightarrow $.

\subsection{Orthogonality of matrix $Q_a$}

\textit{Proposition A1}. $Q_a$ is orthogonal matrix:
\begin{equation}	\label{Eq65}
	{Q_a}Q_a^T = Q_a^T{Q_a} = I.
\end{equation}

\textit{Proof}. (\ref{Eq64}), (\ref{Eq35}), (\ref{Eq32})$\quad \Rightarrow $
\begin{eqnarray*}
	({Q_a} && Q_a^T)_{kl}\operatorname{Tr}\left( {I_a} \right) = {Q_{kia}}{Q_{lja}}{I_{ij}}\operatorname{Tr}\left( {I_a} \right) \\
		&&= {Q_{kia}}{Q_{lja}}\operatorname{Tr}\left( {\sigma}_{ia}{{\sigma}_{ja}} \right) = \operatorname{Tr}\left[ ({Q_{kia}}{\sigma}_{ia})({Q_{lja}}{{\sigma}_{ja}}) \right] \\
		&&= \operatorname{Tr}\left( {U_a}{{\sigma}_{ka}}U_a^{\dagger}{U_a}{{\sigma}_{la}}U_a^{\dagger} \right) = \operatorname{Tr}\left( {{\sigma}_{ka}}{{\sigma}_{la}} \right) \\
    	&&= {I_{kl}}\operatorname{Tr}\left( {I_a} \right)\quad \Rightarrow \quad {{({Q_a}Q_a^T)}_{kl}} = {I_{kl}} \quad \Rightarrow \quad \text{(\ref{Eq65})}. \hspace{6.8pt} \blacksquare
\end{eqnarray*}

\vspace{10pt}
\subsection{Expression for matrix $D_a^T$}

\textit{Proposition A2}. In terms of operators ${\{{L_{na}}\}}_n$ the matrix $D_a^T$ is written as
\begin{equation}	\label{Eq66}
	D_{ija}^T{{\sigma}_{ja}} = \sum\nolimits_n{{L_{na}}{\sigma}_{ia}L_{na}^{\dagger}}.
\end{equation}

\textit{Proof}. (\ref{Eq64}), (\ref{Eq45})$\quad \Rightarrow $
\begin{eqnarray*}
	({D_a} && D_a^T)_{kl}\operatorname{Tr}\left( {I_a} \right) = {D_{kia}}{D_{lja}}{I_{ij}}\operatorname{Tr}\left( {I_a} \right) \\
		&&= {D_{kia}}{D_{lja}}\operatorname{Tr}\left( {\sigma}_{ia}{{\sigma}_{ja}} \right) = {D_{kia}}\operatorname{Tr}\left( {\sigma}_{ia}{D_{lja}}{{\sigma}_{ja}} \right) \\
		&&= {D_{kia}}\operatorname{Tr}\left( {\sigma}_{ia}\sum\nolimits_n{L_{na}^{\dagger}{{\sigma}_{la}}{L_{na}}} \right) \\
		&&= {D_{kia}}\sum\nolimits_n{\operatorname{Tr}\left( {L_{na}}{\sigma}_{ia}L_{na}^{\dagger}{{\sigma}_{la}} \right)} \\
		&&= :{D_{kia}}\operatorname{Tr}\left( {X_{ija}}{{\sigma}_{ja}}{{\sigma}_{la}} \right) = {D_{kia}}{X_{ija}}\operatorname{Tr}\left( {{\sigma}_{ja}}{{\sigma}_{la}} \right) \\
		&&= {{\left( {D_a}{X_a} \right)}_{kl}}\operatorname{Tr}\left( {I_a} \right)\quad \Rightarrow \quad \text{(\ref{Eq66})}. \hspace{79.8pt} \blacksquare
\end{eqnarray*}

\vspace{-11pt}
\subsection{Commutator of matrices $D_a$ and $D_a^T$}

\textit{Proposition A3}. Matrices $D_a$ and $D_a^T$ commute:
\begin{equation}	\label{Eq67}
	[{D_a},D_a^T] = 0.
\end{equation}

\textit{Proof}. (\ref{Eq66}), (\ref{Eq64}), (\ref{Eq45}), (\ref{Eq42})$\quad \Rightarrow $
\begin{eqnarray*}
	(D_a^T && {D_a})_{kl}\operatorname{Tr}\left( {I_a} \right) = D_{kia}^T{D_{jla}}{I_{ij}}\operatorname{Tr}\left( {I_a} \right) \\
		&&= D_{kia}^T{D_{jla}}\operatorname{Tr}\left( {\sigma}_{ia}{{\sigma}_{ja}} \right) = \operatorname{Tr}\left[ (D_{kia}^T{\sigma}_{ia})(D_{lja}^T{{\sigma}_{ja}}) \right] \\
		&&= \sum\nolimits_{mn}{\operatorname{Tr}\left( {L_{ma}}{{\sigma}_{ka}}L_{ma}^{\dagger}{L_{na}}{{\sigma}_{la}}L_{na}^{\dagger} \right)} \\
		&&= \sum\nolimits_{mn}{\operatorname{Tr}\left( L_{na}^{\dagger}{{\sigma}_{ka}}{L_{na}}L_{ma}^{\dagger}{{\sigma}_{la}}{L_{ma}} \right)} \\
		&&= \operatorname{Tr}\left( {D_{kia}}{\sigma}_{ia}{D_{lja}}{{\sigma}_{ja}} \right) = {D_{kia}}D_{jla}^T\operatorname{Tr}\left( {\sigma}_{ia}{{\sigma}_{ja}} \right) \\
		&&= {{({D_a}D_a^T)}_{kl}}\operatorname{Tr}\left( {I_a} \right)\quad \Rightarrow \quad \text{(\ref{Eq67})}. \hspace{76.3pt} \blacksquare
\end{eqnarray*}

\begin{widetext}
\subsection{The contractive property of matrix $D_a$}

\textit{Proposition A4}. In case of POVMs matrix $D_a$ is contractive matrix:
\begin{equation}	\label{Eq68}
	D_a^T{D_a}\le I\quad (\Leftrightarrow \quad \forall b\in {{\mathbb{R}}^3}\quad {b^T}D_a^T{D_a}b\le {b^T}b).
\end{equation}

\textit{Proof}. (\ref{Eq66}), (\ref{Eq64}), (\ref{Eq45}), (\ref{Eq42})$\quad \Rightarrow $
\begin{eqnarray*}
	&&\left. \begin{aligned}
		{b^T}D_a^T{D_a}b\operatorname{Tr}\left( {I_a} \right) &= {b_k}D_{kia}^T{D_{jla}}{b_l}{I_{ij}}\operatorname{Tr}\left( {I_a} \right) = {b_k}D_{kia}^T{D_{jla}}{b_l}\operatorname{Tr}\left( {\sigma}_{ia}{{\sigma}_{ja}} \right) \\
			&= {b_k}{b_l}\operatorname{Tr}\left[ (D_{kia}^T{\sigma}_{ia})(D_{lja}^T{{\sigma}_{ja}}) \right] = {b_k}{b_l}\sum\nolimits_{mn}{\operatorname{Tr}\left( {L_{ma}}{{\sigma}_{ka}}L_{ma}^{\dagger}{L_{na}}{{\sigma}_{la}}L_{na}^{\dagger} \right)} \\
			&= {b_k}{b_l}\frac{1}{2}\sum\nolimits_{mn}{\left[ \operatorname{Tr}\left( {{\sigma}_{ka}}L_{ma}^{\dagger}{L_{na}}{{\sigma}_{la}}L_{na}^{\dagger}{L_{ma}} \right)+\operatorname{Tr}\left( L_{ma}^{\dagger}{L_{na}}{{\sigma}_{la}}L_{na}^{\dagger}{L_{ma}}{{\sigma}_{ka}} \right) \right]} \\
			&= {b_k}{b_l}\frac{1}{2}\sum\nolimits_{mn}{\left\{ \operatorname{Tr}\left[ ({{\sigma}_{ka}}L_{ma}^{\dagger}{L_{na}}){{(L_{ma}^{\dagger}{L_{na}}{{\sigma}_{la}})}^{\dagger}} \right]+\operatorname{Tr}\left[ (L_{ma}^{\dagger}{L_{na}}{{\sigma}_{la}}){{({{\sigma}_{ka}}L_{ma}^{\dagger}{L_{na}})}^{\dagger}} \right] \right\}} \\
			&= \frac{1}{2}\sum\nolimits_{mn}{\left\{ \operatorname{Tr}\left[ ({b_k}{{\sigma}_{ka}}L_{ma}^{\dagger}{L_{na}}){{(L_{ma}^{\dagger}{L_{na}}{{\sigma}_{la}}{b_l})}^{\dagger}} \right]+\operatorname{Tr}\left[ (L_{ma}^{\dagger}{L_{na}}{{\sigma}_{la}}{b_l}){{({b_k}{{\sigma}_{ka}}L_{ma}^{\dagger}{L_{na}})}^{\dagger}} \right] \right\}}, \quad \\
		A	&:= {b_k}{{\sigma}_{ka}}L_{ma}^{\dagger}{L_{na}},\quad B := L_{ma}^{\dagger}{L_{na}}{{\sigma}_{la}}{b_l}, \\
		0	&\le \operatorname{Tr}\left[ (A-B){{(A-B)}^{\dagger}} \right] \quad \Rightarrow \quad \operatorname{Tr}\left( A{B^{\dagger}} \right)+\operatorname{Tr}\left( B{A^{\dagger}} \right)\le \operatorname{Tr}\left( A{A^{\dagger}} \right)+\operatorname{Tr}\left( B{B^{\dagger}} \right)
	\end{aligned} \right\}\Rightarrow \\
	&&\left. \begin{aligned}
		{b^T}D_a^T{D_a}b\operatorname{Tr}\left( {I_a} \right) &\le \frac{1}{2}\sum\nolimits_{mn}{\left\{ \operatorname{Tr}\left[ ({b_k}{{\sigma}_{ka}}L_{ma}^{\dagger}{L_{na}}){{({b_i}{\sigma}_{ia}L_{ma}^{\dagger}{L_{na}})}^{\dagger}} \right]+\operatorname{Tr}\left[ (L_{ma}^{\dagger}{L_{na}}{{\sigma}_{la}}{b_l}){{(L_{ma}^{\dagger}{L_{na}}{{\sigma}_{ja}}{b_j})}^{\dagger}} \right] \right\}} \\
			&= \frac{1}{2}\sum\nolimits_{mn}{\left[ {b_k}{b_i}\operatorname{Tr}\left( {{\sigma}_{ka}}L_{ma}^{\dagger}{L_{na}}L_{na}^{\dagger}{L_{ma}}{\sigma}_{ia} \right)+{b_l}{b_j}\operatorname{Tr}\left( L_{ma}^{\dagger}{L_{na}}{{\sigma}_{la}}{{\sigma}_{ja}}L_{na}^{\dagger}{L_{ma}} \right) \right]} \\
			&= \frac{1}{2}\sum\nolimits_{mn}{\left[ {b_k}{b_i}\operatorname{Tr}\left( {{\sigma}_{ka}}{L_{ma}}L_{ma}^{\dagger}{L_{na}}L_{na}^{\dagger}{\sigma}_{ia} \right)+{b_l}{b_j}\operatorname{Tr}\left( {L_{ma}}L_{ma}^{\dagger}{L_{na}}L_{na}^{\dagger}{{\sigma}_{la}}{{\sigma}_{ja}} \right) \right]} \\
			&= \frac{1}{2}\left[ {b_k}{b_i}\operatorname{Tr}\left( {{\sigma}_{ka}}{\sigma}_{ia} \right)+{b_l}{b_j}\operatorname{Tr}\left( {{\sigma}_{la}}{{\sigma}_{ja}} \right) \right] = \frac{1}{2}\left[ {b_k}{b_i}{I_{ki}}\operatorname{Tr}\left( {I_a} \right)+{b_l}{b_j}{I_{lj}}\operatorname{Tr}\left( {I_a} \right) \right] \\
			&= \frac{1}{2}\left[ {b_i}{b_i}\operatorname{Tr}\left( {I_a} \right)+{b_j}{b_j}\operatorname{Tr}\left( {I_a} \right) \right] = {b_i}{b_i}\operatorname{Tr}\left( {I_a} \right)\quad \Rightarrow \quad \text{(\ref{Eq68})}. \hspace{174.2pt} \blacksquare \\
	\end{aligned} \right.
\end{eqnarray*}
\end{widetext}

\section{eigenvalues of operator ${\sigma}_{i1}{\sigma}_{j2}$}\label{EigVal}

In this appendix, we will find eigenvalues of the operator ${\sigma}_{i1}{\sigma}_{j2}$ which were used in the section \ref{TensExp} at the description of procedure of experiment for measurement of spin correlation tensor (\ref{Eq04}).

\textit{Proposition B1}. The operator ${\sigma}_{i1}{\sigma}_{j2}$ has twice degenerate eigenvalues which are equal to $\pm 1$.

\textit{Proof}. From definition (\ref{Eq07}) we have
\begin{eqnarray*}
	{\sigma}_{i1}{\sigma}_{j2} =&& \; \left( {{| + \rangle}_1}\sigma _i^{11}{}_1\langle + |+{{| + \rangle}_1}\sigma _i^{12}{}_1\langle - | \right. \\
		&& \ \ \left. +{{| - \rangle}_1}\sigma _i^{21}{}_1\langle + |+{{| - \rangle}_1}\sigma _i^{22}{}_1\langle - | \right) \\
		&& \times \left( {{| + \rangle}_2}\sigma _j^{11}{}_2\langle + |+{{| + \rangle}_2}\sigma _j^{12}{}_2\langle - | \right. \\
		&& \quad \ \left. +{{| - \rangle}_2}\sigma _j^{21}{}_2\langle + |+{{| - \rangle}_2}\sigma _j^{22}{}_2\langle - | \right) = ... \\
	=&& \; \left[ \begin{matrix}
			{{| + \rangle}_1}{{| + \rangle}_2} & {{| - \rangle}_1}{{| + \rangle}_2} & {{| + \rangle}_1}{{| - \rangle}_2} & {{| - \rangle}_1}{{| - \rangle}_2} \\
		\end{matrix} \right] \\
		&& \times \left[ \begin{matrix}
			\sigma _i^{11}\sigma _j^{11} & \sigma _i^{12}\sigma _j^{11} & \sigma _i^{11}\sigma _j^{12} & \sigma _i^{12}\sigma _j^{12} \\
			\sigma _i^{21}\sigma _j^{11} & \sigma _i^{22}\sigma _j^{11} & \sigma _i^{21}\sigma _j^{12} & \sigma _i^{22}\sigma _j^{12} \\
			\sigma _i^{11}\sigma _j^{21} & \sigma _i^{12}\sigma _j^{21} & \sigma _i^{11}\sigma _j^{22} & \sigma _i^{12}\sigma _j^{22} \\
			\sigma _i^{21}\sigma _j^{21} & \sigma _i^{22}\sigma _j^{21} & \sigma _i^{21}\sigma _j^{22} & \sigma _i^{22}\sigma _j^{22} \\
		\end{matrix} \right] \! \! \left[ \begin{matrix}
			{}_2\langle + |{}_1\langle + | \\
			{}_2\langle + |{}_1\langle - | \\
			{}_2\langle - |{}_1\langle + | \\
			{}_2\langle - |{}_1\langle - | \\
		\end{matrix} \right] \! .
\end{eqnarray*}

Let's solve a problem on eigenvalues for the obtained $4\times 4$ matrix:
\begin{equation}	\label{Eq69}
	\left[ \begin{matrix}
		{{\sigma}_i}\sigma _j^{11} & {{\sigma}_i}\sigma _j^{12} \\
		{{\sigma}_i}\sigma _j^{21} & {{\sigma}_i}\sigma _j^{22} \\
	\end{matrix} \right]
	\left[ \begin{matrix} u \\ v \\ \end{matrix} \right]
	= \lambda
	\left[ \begin{matrix} u \\ v \\ \end{matrix} \right].
\end{equation}
For the solution of the equation on eigenvalues, we use properties of determinant:
\begin{equation}	\label{Eq70}
	\begin{aligned}
		\det \left[ \begin{matrix}
			A & B \\
			C & D \\
		\end{matrix} \right]
		&= \det \left( A \right)\det \left( D-C{A^{-1}}B \right), \\
		\det \left( cA \right)
		&= {c^n}\det \left( A \right)\quad (A\in {{\mathbb{C}}^{n\times n}}).
	\end{aligned}
\end{equation}
At various values of an index $j$ taking into account expressions (\ref{Eq08}), (\ref{Eq69}) and (\ref{Eq70}) we write
\begin{eqnarray*}
	&&\left. \begin{aligned}
	j = x & \quad \Rightarrow \quad 0 = \left| \begin{matrix}
			-\lambda I & {{\sigma}_i} \\
			{{\sigma}_i} & -\lambda I \\
		\end{matrix} \right| \\
		&= \det \left( -\lambda I \right) \det \left( -\lambda I + {{\sigma}_i}{{\lambda}^{-1}}I{{\sigma}_{\underline{i}}} \right) \quad \Rightarrow ... \quad \\
	j = y & \quad \Rightarrow \quad 0 = \left| \begin{matrix}
			-\lambda I & -i{{\sigma}_i} \\
			i{{\sigma}_i} & -\lambda I \\
		\end{matrix} \right| \\
		&= \det \left( -\lambda I \right) \det \left( -\lambda I + {{\sigma}_i}{{\lambda}^{-1}}I{{\sigma}_{\underline{i}}} \right) \quad \Rightarrow ... \\
	j = z & \quad \Rightarrow \quad 0 = \left| \begin{matrix}
			{{\sigma}_i} - \lambda I & O \\
			O & -{{\sigma}_i} - \lambda I \\
		\end{matrix} \right| \\
		&= \det \left({{\sigma}_i} - \lambda I \right) \det \left(-{{\sigma}_{\underline{i}}}-\lambda I \right) \quad \Rightarrow ... \\
	\end{aligned} \right] \Rightarrow \\
	&&\left. \begin{aligned}
	&{{\lambda}_{1,2}} = +1, \quad {{\lambda}_{3,4}} = -1 .
	\end{aligned} \right.
\end{eqnarray*}
Here $O$ is zero matrix, on the repeating index with underlining, there is no summing up. Thus, the operator ${\sigma}_{i1}{\sigma}_{j2}$ has twice degenerate eigenvalues which are equal to~$\pm 1$.~\hfill$\blacksquare$

\bibliography{Article}

\end{document}